\documentclass[preprint2]{aastex}  

\slugcomment{Submitted to the Astrophysical Journal}

\shorttitle{Weak Lensing from Space III}
\shortauthors{Refregier et al.}

\begin{document}

\title{Weak Lensing from Space III: \\Cosmological Parameters}

\author{Alexandre Refregier\altaffilmark{1,2}}
\affil{Service d'Astrophysique, CEA Saclay,
        91191 Gif sur Yvette, France}
\email{refregier@cea.fr}

\author{Richard Massey}
\affil{Institute of Astronomy, Madingley Road, Cambridge CB3 OHA, U.K.}

\author{Jason Rhodes\altaffilmark{3,4}}
\affil{California Institute of Technology, 1201 E. California Blvd,
        Pasadena, CA 91125}
\and

\author{Richard Ellis\altaffilmark{2},
        Justin Albert\altaffilmark{2},
        David Bacon\altaffilmark{5}, \\
        Gary Bernstein\altaffilmark{6},
        Tim McKay\altaffilmark{7} \& 
        Saul Perlmutter\altaffilmark{8}}

\altaffiltext{1}{Institute of Astronomy, Madingley Road, Cambridge CB3 OHA,
         U.K.}
\altaffiltext{2}{California Institute of Technology, 1201 E. California Blvd,
         Pasadena, CA 91125}
\altaffiltext{3}{Laboratory for Astronomy \& Solar Physics, Code 681,
         Goddard Space Flight Center, Greenbelt, MD 20771}
\altaffiltext{4}{NASA/NRC Research Associate}
\altaffiltext{5}{Institute for Astronomy, Blackford Hill, Edinburgh EH9
         3HJ, U.K.}
\altaffiltext{6}{Department of Physics \& Astronomy, Univ. of Pennsylvania,
         209 S. 33$^{\rm rd}$ Street, Philadelphia, PA 19104}
\altaffiltext{7}{Department of Astronomy, University of Michigan, Ann
         Arbor, MI 48109}
\altaffiltext{8}{Lawrence Berkeley National Laboratory, 1 Cyclotron Road,
         Berkeley, CA 94720}

\begin{abstract}
Weak gravitational lensing provides a unique method to directly map
the dark matter in the universe and measure cosmological
parameters. Current weak lensing surveys are limited by the
atmospheric seeing from the ground and by the small field of view of
existing space telescopes. We study how a future wide-field space
telescope can measure the lensing power spectrum and skewness, and set
constraints on cosmological parameters. The lensing sensitivity was
calculated using detailed image simulations and instrumental
specifications studied in earlier papers in this series. For instance,
the planned \emph{SuperNova/Acceleration Probe} (SNAP) mission will be
able to measure the matter density parameter $\Omega_{m}$ and the dark
energy equation of state parameter $w$ with precisions comparable and
nearly orthogonal to those derived with SNAP from supernovae. The
constraints degrade by a factor of about 2 if redshift tomography is
not used, but are little affected if the skewness only is dropped. We
also study how the constraints on these parameters depend on the
survey geometry and define an optimal observing strategy.
\end{abstract}


\section{Introduction}
Weak gravitational lensing provides a unique method to directly
map the distribution of mass in the universe (for reviews see
Bartelmann \& Schneider 2001; Mellier et al. 2002; Hoekstra et al.
2002; Refregier 2003). The coherent distortions that lensing
induces on the shape of background galaxies have now been firmly
measured from the ground and from space. The amplitude and angular
dependence of this `cosmic shear' signal can be used to set strong
constraints on cosmological parameters. Several surveys are now in
progress to map larger areas and thus reduce the uncertainties in
these parameters. However, future ground based surveys will
eventually be limited by the systematics induced by atmospheric
seeing. Space based observations do not suffer from this effect,
but their statistics are currently  limited by the small field of
view of existing space telescopes.

In this paper series, we study how these limitations can be
circumvented using wide field imaging from space, using the planned
\emph{SuperNova/Acceleration Probe} (SNAP) mission (Perlmutter et
al. 2003) as a concrete example. In the first paper in this series
(Rhodes et al. 2003; Paper I), we study the instrumental
characteristics and survey stategy of such a mission, showing that it
would provide both excellent statistics and reduced systematics
relevant for weak lensing. In a subsequent paper (Massey et al. 2003;
Paper II), we used detailed image simulations to compute the
sensitivity for measuring the weak lensing shear from space, and thus
to derive high resolution maps of the Dark Matter in the local
universe.

In this paper, we use the previously derived lensing sensitivity (see
Papers I and II) to determine the constraints that can be placed on
cosmological parameters via weak lensing from space. We consider
quintessence (QCDM) models with a dark energy component with arbitrary
constant equation of state parameter $w$. We compute the lensing power
spectrum and skewness and their associated errors for different survey
strategies. We study how the photometric redshifts derived from the
SNAP filter set can be used to study the evolution of the lensing
power spectrum. We then compare the resulting lensing constraints
on cosmological parameters with those derived from supernovae.
Earlier studies of the constraints on dark energy from generic weak
lensing surveys can be found in Hui (1999), Huterer (2001), Benabed \&
Bernardeau (2001), Hu (2001), Weinberg \& Kamionkowski (2002), Munshi
\& Wang (2002).  While these authors have considered generic weak
lensing surveys, we use realistic redshift distributions, lensing
sensitivities, and photometric-redshift errors relevant for the
concrete case of SNAP. This allows us to include the effects of
photometric-redshift errors and leakage between redshift bins, and to
study the trade off between width and depth in future surveys. We also
study how the measurement of the skewness can be combined with power
spectrum tomography to improve the accuracy of the determination of
cosmological parameters.

This paper is organized as follows. In \S\ref{snap}, we summarize
the characteristics of the SNAP mission. In \S\ref{photoz}, we
describe its capabilities for deriving photometric redshifts. In
\S\ref{cosmology}, we describe the cosmological models we will
consider. In \S\ref{power}, we compute the lensing power spectrum,
its associated errors, and its redshift evolution. In
\S\ref{skewness}, we compute the skewness of the shear field and
associated errors. In \S\ref{parameters}, we compute the
constraints which can be set on cosmological parameters from
measurements of the power spectrum and skewness.  Our conclusions
are summarized in \S\ref{conclusion}.

\section{The SNAP Mission}
\label{snap} The SNAP satellite will consist of a 2 meter
telescope in space with a field of view of 0.7 deg$^{2}$ (Perlmutter
et al. 2003; see also Paper I). The mission lifetime will be divided
between two deep 16 months survey and a 5 month wide survey. The deep
surveys will cover 15 deg$^{2}$ and are primarily designed for the
search for Type Ia supernovae. It will also be invaluable to map the
dark matter via weak lensing (see Paper II). The wide survey is
designed primarily for weak lensing and will cover 300 deg$^{2}$. The
spacecraft will be in a high elliptical orbit with good thermal
stability, thus affording stable image quality and a low level of
systematics. The details of the performance of the instrument for weak
lensing and of the survey strategy can be found in Paper I.

\begin{table*}
 \begin{center}
  \caption{Survey parameters and redshift distributions\tablenotemark{a} \label{tab:nz}}
  \begin{tabular}{lclclllllllllll}
   \tableline \tableline 
   Survey & $z$-bins & $t_{\rm exp}$\tablenotemark{b} & $t_{\rm tot}$ & 
   $A$ & $n_{g}$ & $\sigma_\gamma$\tablenotemark{c} & $z_{m}$ & $z_{0}$ &
   $\alpha$ & $\beta$ & $z_{+}$ & $\zeta_{+}$ & $z_{-}$ & $\zeta_{-}$ \\
      &     & (sec) & (months) & (deg$^{2}$) & (amin$^{-2}$) & \\
  \tableline
Deep  &     & 20000 & 32 &  15 & 260 & 0.36 & 1.43 & 1.31  & 2.00 & 2.00 \\
Wide  &     & 2000  &  5 & 300 & 100 & 0.31 & 1.23 & 1.13  & 2.00 & 2.00 \\
      & 1/2 & 2000  &  5 & 300 & 50  & 0.31 & 0.96 & 1.32  & 1.94 & 3.38 & 1.36 & 0.042 &      &       \\
      & 2/2 & 2000  &  5 & 300 & 50  & 0.31 & 1.73 & 1.51  & 0.53 & 2.16 &      &       & 1.36 & 0.048 \\
      & 1/3 & 2000  &  5 & 300 & 33  & 0.31 & 0.81 & 1.13  & 1.95 & 5.55 & 1.11 & 0.031 &      &       \\
      & 2/3 & 2000  &  5 & 300 & 33  & 0.31 & 1.31 & 0.80  &20.07 & 3.45 & 1.11 & 1.515 & 1.59 & 1.515 \\
      & 3/3 & 2000  &  5 & 300 & 33  & 0.31 & 1.93 & 1.57  & 1.50 & 2.48 &      &       & 1.59 & 0.042 \\
Wide+ &     & 1000  &  5 & 600 & 68  & 0.30 & 1.17 & 1.07  & 2.00 & 2.00 \\
Wide$-$ &   & 4000  &  5 & 150 & 150 & 0.33 & 1.31 & 1.20  & 2.00 & 2.00 \\
   \tableline
  \end{tabular}
  \tablenotetext{a}{The redshift bin distributions assume the use of the 9 SNAP filters, including the Near--IR detectors.}
  \tablenotetext{b}{Exposure time in each optical filter, equal to half of the exposure time for the near--IR filters.}
  \tablenotetext{c}{rms shear $\sigma_{\gamma}=\langle|\gamma|^{2}\rangle^{\frac{1}{2}}$ from noise and intrinsic ellipticity.}
 \end{center}
\end{table*}

\section{Photometric Redshifts}
\label{photoz} The SNAP focal plane will be partially covered by
CCDs sensitive to 9 optical and near--IR bands. Paper II describes how
this filter set affords excellent photometric redshifts. This was
tested using the {\tt HyperZ} code (Bolzonella et al. 2000) to
generate simulated galaxy spectra and recovered photometric
redshifts. Using all nine filters, we found that redshifts can be
recovered with a $1\sigma$ precision better than 0.03. Including the
near--IR detectors prevents catastrophic failures in redshift
estimation by eliminating strong degeneracies between low ($z \lesssim
0.5$) and high ($z \gtrsim 1$) redshift bins (see Paper II).

\begin{figure}[htbp]
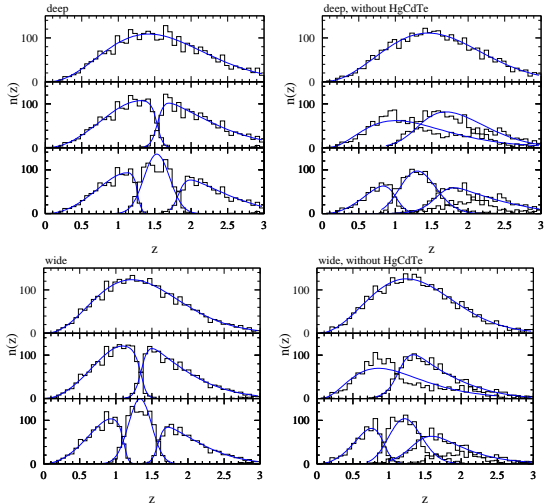

 \epsscale{1}
 \plottwo{Refregier.fig1a.epsi}{Refregier.fig1b.epsi} 
 \epsscale{2.2}
 \plottwo{Refregier.fig1c.epsi}{Refregier.fig1d.epsi}
 \epsscale{0.7}
 \caption{Redshift bins derived from photomotric redshifts. In each of the 4
 panels, the histograms show the redshift distributions resulting from cuts in
 photometric redshifts aimed to produce 1,2 and 3 redshift bins from top to
 bottom, respectively. The solid curves correspond to fits for the analytical
 form of equations~(\ref{eq:nz}-\ref{eq:f_pm}). The top and bottom panels
 correspond to the deep and wide SNAP surveys respectively. In the left panels,
 the full set of 9 optical and near--IR (HgCdTe) SNAP filters were used to
 estimate the photometric redshifts. In the right panels, only the 6 optical
 filters were used. In all cases, the normaliszation is arbitrary.}
 \label{fig:zbins}
\end{figure}

These high precision photometric redshifts will allow us to
construct 3-dimensional maps of the dark matter (see Paper II).
They will also be useful to study the evolution of the lensing
statistics. Figure~\ref{fig:zbins} shows how photometric redshifts
can be used to group galaxies into redshift bins. The input
redshift distribution $n(z)$ was assumed to have the form
\begin{equation}
\label{eq:nz}
n(z) \propto z^{\alpha} e^{-(z/z_{0})^{\beta}}
\end{equation}
where $z_{0}$, $\alpha$ and $\beta$ are parameters estimated from
existing deep redshift surveys (see Paper II).  Table~\ref{tab:nz}
lists the values of these parameters for the deep and wide
surveys, along with the associated median redshift $z_{m}$, the
surface density $n_{g}$ of galaxies usable for lensing, and the
survey solid angle $A$. The exposure time $t_{\rm exp}$ for each
optical filter, along with the total observing $t_{\rm obs}$ time
for the survey are also listed. The figure shows the redshift
distribution resulting from binning the galaxies into 2 and 3
photometric-redshift bins, with approximately the same number of
galaxies in each bin. With the near--IR detectors (left panels),
the photometric redshifts afford excellent separation between the
bins.  In the absence of these detectors (right panels), the
leakage between bins degrades, due to the increased noise and
degeneracies in the photometric redshifts (see Figure 7 in Paper
II). In the following, we will always assume that the near--IR
detectors are available.

The redshift distributions of each bin can be described analytically
by multiplying the input redshift distribution $n(z)$ in
equation~(\ref{eq:nz}) by the high-$z$ and low-$z$ filter functions
$f_{+}(z)$ and/or $f_{-}(z)$ given by
\begin{equation}
\label{eq:f_pm}
f_{\pm}(z) = \left[ 1+ e^{\pm(z_{\pm}-z)/\zeta_{\pm}} \right]^{-1},
\end{equation}
where $z_{+}$ and $z_{-}$ are the cut off redshifts, and
$\zeta_{+}$ and $\zeta_{-}$ are smearing factors arising from the
finite photometric redshift accuracy. Fits to the redshift bin
distributions using these analytical forms are shown in
Figure~\ref{fig:zbins}.  The values of the resulting parameters
for 2 and 3 redshift bins are listed in Table~\ref{tab:nz}.  In
\S\ref{tomography}, we study how multiple redshift bins can be
used to measure the evolution of the lensing power spectrum and
thus improve the accuracy of the measurement of cosmological
parameters.

\section{Cosmological Model}
\label{cosmology} We consider a cosmology with an expansion
parameter $a=(1+z)^{-1}$ that is determined by a matter component and
a dark energy (or `quintessence') component with present density
parameters $\Omega_{m}$ and $\Omega_{q}$, respectively.  The equation
of state of the dark energy is parametrized by $w=p_{q}/\rho_{q}$,
which we assume to be constant and is equal to -1 in the case of a
cosmological constant. The evolution of the expansion parameter is
given by the Hubble constant $H$ through the Friedmann equation
\begin{equation}
H=\frac{\dot{a}}{a}=H_{0}\left( \Omega_{m} a^{-3} + \Omega_{q}
a^{-3(1+w)} + \Omega_{\kappa} a^{-2} \right)^{\frac{1}{2}},
\end{equation}
where $\dot{a}=da/dt$ and the total and curvature density
parameters are $\Omega$ and $\Omega_{\kappa}=1-\Omega$,
respectively. The present value of the Hubble constant is
parametrized as $H_{0}=100 h$ km s$^{-1}$ Mpc$^{-1}$.

As a reference model, we consider a fiducial $\Lambda$CDM model with
parameters $\Omega_{m}=0.30$, $\Omega_{b}=0.047$, $n=1$, $h=0.7$,
$w=-1$, consistent with the recent WMAP experiment (see tables 1-2 in
Spergel et al. 2003). In agreement with this experiment, we assume
that the universe is flat, i.e. that $\Omega=\Omega_{m}+\Omega_{q}=1$.
(Note that, in our notation, $\Omega_{m}$ includes both dark matter
and baryons). The shape parameter for the matter power spectrum is
taken to be $\Gamma=\Omega_{m}h
\exp{[-\Omega_{b}(1+\sqrt{2h}/\Omega_{m})]}$ as prescribed by Sugiyama
(1995). The matter power spectrum is normalized according to the COBE
normalization (Bunn \& White 1996), which corresponds to
$\sigma_{8}=0.88$. This is consistent with the WMAP results (Spergel
et al. 2003) and with the average of recent cosmic shear measurements
(see compilation tables in Mellier et al. 2002; Hoekstra et al. 2002;
Refregier et al. 2003). In the following, we will consider deviations
from this reference model.

\section{Weak Lensing Power Spectrum}
\label{power}

\subsection{Theory}
The weak lensing power spectrum is given by (eg. Bartelmann \&
Schneider 1999; Hu \& Tegmark 1999; see Bacon et al. 2001 for
conventions)
\begin{equation}
\label{eq:cl}
C_{\ell} = \frac{9}{16} \left( \frac{H_{0}}{c}
\right)^{4} \Omega_{m}^{2}
  \int_{0}^{\chi_h} d\chi~\left[ \frac{g(\chi)}{a r(\chi)} \right]^{2}
  P\left(\frac{\ell}{r}, \chi\right),
\end{equation}
where $r(\chi)$ is the comoving angular diameter distance, and
$\chi_{h}$ corresponds to the comoving radius to the horizon. The
non-linear matter power spectrum $P(k,z)$ is computed using the
transfer function from Bardeen et al. (1986; with the conventions of
Peacock 1997), thus ignoring the corrections on large scales for
quintessence models (Ma et al. 1999). The growth factor and COBE
normalization for arbitrary values of $w$ was computed using the
fitting formulae from Ma et al. (1999). Considerable uncertainties
remain for the non-linear corrections in quintessence models (see
discussion in Huterer 2001). Here, we use the fitting formula from
Peacock \& Dodds (1996) but acknowledge that it differs significantly
from that from Ma et al. (1999). The impact of this uncertainty is
discussed below in \S\ref{conclusion}. \label{pd}

The radial weight function $g$ is given by
\begin{equation}
g(\chi) = 2 \int_{\chi}^{\chi_{h}} d\chi'~n(\chi')
   \frac{r(\chi)r(\chi'-\chi)}{r(\chi')},
\end{equation}
where $n(\chi)$ is the probability of finding a galaxy at comoving
distance $\chi$ and is normalised as $\int d\chi~n(\chi) =1$. For our
purposes, we use the analytical fits for $n(z)$ given in
equations~(\ref{eq:nz}-\ref{eq:f_pm}) along with the parameter values
listed in table~\ref{tab:nz}.

\begin{figure}
 \plotone{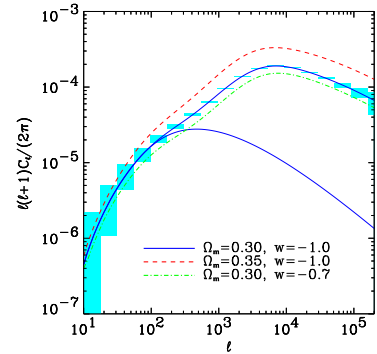}
 \caption{Weak lensing power spectrum for several cosmological
 models. The top solid line shows the weak lensing power spectrum
 $C_{\ell}$ for the fiducial $\Lambda$CDM model with
 $\Omega_{m}=0.30$, and $w=-1$. The bottom solid line shows the
 linear power spectrum for the same model. The dashed and
 dot-dashed lines show the nonlinear power spectra for variations
 of the model with $\Omega_{m}=0.35$ and $w=-.7$ respectively. In
 all cases, $\Omega_{q}=1-\Omega_{m}$, $h=0.7$, $\Omega_{b}=0.047$,
 $n=1$ and COBE normalization were assumed. The redshift
 distribution was taken to be that for the SNAP wide survey
 (unbinned) with a median redshift of $z_m=1.23$.  The boxes
 correspond to the band averaged $1\sigma$ errors about the
 fiducial model for the SNAP wide survey (300 deg$^{2}$ area, 100
 galaxies per arcmin$^{2}$ and an intrinsic shear dispersion of
 $\sigma_{\gamma}=0.31$).
 \label{fig:cls_omw}}
\end{figure}

Figure~\ref{fig:cls_omw} shows the lensing power spectrum for the
fiducial $\Lambda$CDM model. Deviations from the model corresponding
to variations in $\Omega_{m}$ and $w$ are also shown. All models shown
are COBE normalised.  The linear power spectrum for the fiducial model
is also shown, highlighting the importance of non-linear evolution for
$\ell \gtrsim 100$.

\subsection{Measurement uncertainties}
Neglecting non-gaussian corrections, the rms uncertainty in measuring
the lensing power spectrum $C_{\ell}$ is given by (Kaiser 1998; Hu \&
Tegmark 1999; Huterer 2001)
\begin{equation}
\Delta C_{l}=\sqrt{ \frac{2}{(2l+1) f_{\rm
sky}}}\left(C_{l}+\frac{\sigma_{\gamma}^{2}}{2n_{g}} \right),
\end{equation}
where $f_{\rm sky}$ is the fraction of the sky covered by the survey,
$n_{g}$ is the surface density of usable galaxies, and
$\sigma_{\gamma}^{2} =\langle |\gamma|^{2} \rangle$ is the shear variance per
galaxy arising from intrinsic shapes and measurement errors. Values of
$\sigma_{\gamma}$ for the different SNAP surveys were derived from the
image simulations in Paper II and are listed in table~\ref{tab:nz}.

\begin{figure}
 \plotone{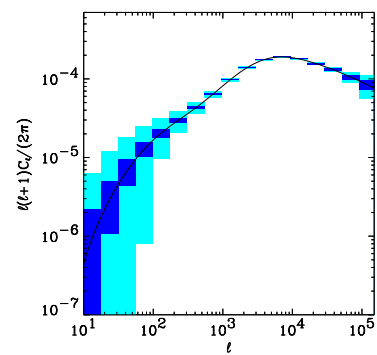}
 \caption{Measurement of the weak lensing power spectrum with the
 wide and deep SNAP surveys. The solid line shows the power spectrum
 for the fiducial $\Lambda$CDM model of the previous figure. The light
 and dark boxes show the band-averaged $1\sigma$ errors for the deep and
 wide surveys respectively.}
 \label{fig:cls_wd}
\end{figure}

Figure~\ref{fig:cls_omw} shows the resulting band averaged errors
for the fiducial $\Lambda$CDM model measured with the SNAP weak
lensing survey. The sensitivity afforded by this survey is
excellent, and will allow us to easily distinguish between the
different cosmological models shown. Figure~\ref{fig:cls_wd}
compares the precision expected for the wide and deep SNAP
surveys. The deep survey clearly yields lower precision for the
measurement of the power spectrum, in spite of its longer observing
time. It will however be ideally suited to produce high resolution
maps of the dark matter (see Paper II).
\label{cls_wd}

\subsection{Evolution of the Power Spectrum}
\label{tomography} As discussed in \S\ref{photoz}, the SNAP filter
set will allow us to divide the galaxies into several redshift
bins. Possible redshift bin configurations are shown in
Figure~\ref{fig:zbins}. The lensing power spectrum can then be
measured  separately in each bin, yielding a tomography of the
mass distribution along the line of sight (Hu 1999; Hu \& Keeton
2002; Taylor 2001).

\begin{figure}
 \plotone{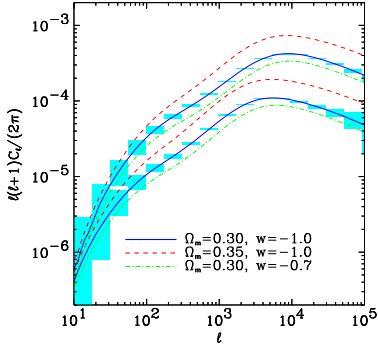}
 \caption{Redshift dependence of the lensing power spectrum.  The
 solid lines and associated $1\sigma$ error boxes show the lensing
 power spectrum for the two redshift bins of the SNAP wide survey
 with median galaxy redshifts of $z_{m}=0.96$ (bottom line) and
 1.73 (top line). As in figure~\ref{fig:cls_omw}, the dashed and
 dot-dashed lines correspond to perturbations about the fiducial
 model (solid line) for each redshift bin.}
 \label{fig:cls_z}
\end{figure}

Figure~\ref{fig:cls_z} shows, for instance, the lensing power spectrum
and associated error bars for the two redshift bins derived from the
SNAP wide survey with median redshifts $z_{m}\simeq0.96$ and $1.73$
(see figure~\ref{fig:zbins} and table~\ref{tab:nz}). Clearly the
amplitude of the power spectrum is much larger for the more distant
bin. The sensitivity afforded by the SNAP wide survey will allow us to
easily measure each power spectrum separately. In
\S\ref{params_tomography} below, we show how the measurement of the
lensing power spectrum at different redshifts improves the precision
of cosmological parameters.

\section{Skewness}
\label{skewness}
Non-linear gravitational instability is known to produce non-gaussian
features in the cosmic shear field. The power spectrum therefore does
not contain all the information available from weak lensing. We
consider the most common measure of non-gaussianity, namely the
skewness $S_{3}$ which is defined as (eg. Bernardeau et al. 1997)
\begin{equation}
\label{eq:s3}
S_{3}(\theta) \equiv \frac{\langle \kappa^{3}
\rangle}{\langle \kappa^2 \rangle^2}
\end{equation}
where $\kappa$ is the convergence which can be derived from the
shear field $\gamma_{i}$ and the brackets denote averages over
circular top-hat cells of radius $\theta$. The denominator is the
square of the convergence variance which is given by
\begin{equation}
\langle \kappa^{2} \rangle = \langle \gamma^{2} \rangle \simeq
\frac{1}{2\pi} \int d\ell~\ell C_{\ell}|W_{\ell}|^{2},
\end{equation}
where $W_{\ell} \equiv 2J_{1}(\ell\theta)/(\ell\theta)$ is the window
function for such cells and $C_{\ell}$ is the lensing power spectrum
given by equation~(\ref{eq:cl}).

To evaluate the numerator of equation~(\ref{eq:s3}) we use the
approximation of Hui (1999) who used the Hyperextended perturbation
theory of Scoccimarro \& Frieman (1999) and obtained
\begin{eqnarray}
\langle \kappa^{3} \rangle & \simeq & \frac{81 \pi^{2}}{16}
\left( \frac{H_{0}}{c} \right)^{6} \Omega_{m}^{3} \times \nonumber \\
& & \int_{0}^{\chi_{h}} d\chi~\frac{g^3}{a^{3}r^4} \left[ \int
d^{2}\ell~\sqrt{Q_{3}} P\left( \frac{\ell}{r},\chi \right) |W_{\ell}|^{2}
\right]^{2},
\end{eqnarray}
where $Q_{3}=(4-2^{n})/(1+2^{n+1})$ and $n$ is the linear power
spectral index at scale $k=\ell/r$. While more accurate
approximations for third order statistics now exist (see van
Waerbeke et al. 2001 and reference therein), the present one
suffices for our purpose.

\begin{figure}
 \plotone{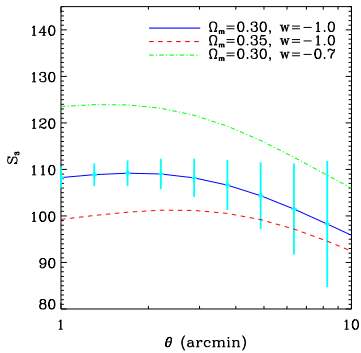}
 \caption{Skewness $S_{3}$ as a function of scale. The three
 cosmological models from Figure~\ref{fig:cls_omw} are displayed.
 The $1\sigma$ error bars correspond to the SNAP wide survey.}
 \label{fig:s3}
\end{figure}

Figure~\ref{fig:s3} shows the skewness as a function of scale for the
same cosmological models considered in Figure~\ref{fig:cls_omw}. The
skewness is only weakly dependent on the angular scale $\theta$, but
depends more strongly on $\Omega_{m}$ and $w$.

The computation of the exact error for $S_{3}$ is challenging as
it depends on sixth order terms, which are difficult to compute in
the non-linear regime. Instead, we compute the rms error for a
gaussian field (in which case $S_{3}=0$) and introduce a
multiplicative factor $f_{ng}$ to correct for non-Gaussianity of
the convergence field and obtain
\begin{equation}
(\Delta S_{3})^{2} = \frac{15}{N_{c}} \frac{ \left[ f_{ng}^\frac{2}{3} \langle
\kappa^{2} \rangle + \sigma_{\kappa}^{2}/(n_{g}A_{c})
\right]^{3} }{\langle \kappa^{2} \rangle^{2}},
\end{equation}
where $A_{c}=\pi \theta^{2}$ is the cell solid angle,
$N_{c}=A/A_{c}$ is the number of cells which are assumed to be
independent, and $A$ is the total solid angle of the survey. The
rms dispersion of the convergence arising from the intrinsic
dispersion of the galaxy ellipticities and from measurement noise
is related to the associated rms shear by
$\sigma_{\kappa}^{2}=\sigma_{\gamma}^2$. The non-gaussian
correction factor only applies to the cosmic variance term (first
term) since the noise term can be assumed to be gaussian. It is
set to $f_{ng}\simeq 2$, as estimated by White \& Hu (2000) who
compared gaussian estimates with errors derived from (noise-free)
numerical simulations.

The resulting errors for SNAP wide survey are shown in
Figure~\ref{fig:s3}, for the fiducial $\Lambda$CDM model. The
sensitivity afforded by this survey will allow us to easily
distinguish between these models via the skewness.

\section{Constraints on Cosmological Parameters}
\label{parameters}

\subsection{Fisher Matrix}
The constraints which can be set on cosmological parameters can be
estimated using the Fisher matrix (eg. Hu \& Tegmark 1999)
\begin{equation}
F_{ij}= - \left\langle \frac{\partial \ln {\mathcal L}}{\partial
p_{i} \partial p_{j}} \right \rangle
\end{equation}
where ${\mathcal L}$ is the Likelihood function, and $p_{i}$ is a
set of model parameters. The inverse ${\mathbf F}^{-1}$ provides a
lower limit for the covariance matrix of the parameters.

For a measurement of the power spectrum this reduces to,
\begin{equation}
F_{ij}= \sum_{\ell} (\Delta
C_{\ell})^{-2} \frac{\partial C_{\ell}}{\partial p_{i}} \frac{\partial
C_{\ell}}{\partial p_{j}},
\end{equation}
where the summation is over modes $\ell$ which can be reliably
measured.  Note that this expression assumes that the errors are
gaussian and that the multipoles are not correlated. These effects
have been shown to increase the errors on cosmological parameters by
only about 15\% (Cooray \& Hu 2001) and have been neglected here.

Since the measurement of the skewness on different scales are
strongly correlated, we conservatively  consider only one scale
$\theta=2'$ to compute the constraints from $S_{3}$. The
associated fisher matrix is then
\begin{equation}
F_{ij}= (\Delta S_{3})^{-2} \frac{\partial^{2} S_{3}}{\partial p_{i}\partial p_{j}}.
\end{equation}
The joint constraints from the power spectrum combined with the
skewness can be computed by adding the respective Fisher
matrices. \label{skew_onescale}

\subsection{Baseline surveys}

\begin{figure}
 \plotone{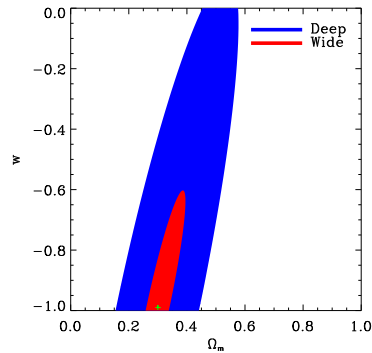}
 \caption{Constraints on $\Omega_{m}$ and $w$ from the power
 spectrum derived from the wide and deep SNAP surveys. The contours
 correspond to the 68\% confidence level and have been marginalized
 over $h$, $n$ and $\Omega_{b}$, with a 7\% rms COBE prior for the
 power spectrum normalization $\delta_{h}$. The cosmological model
 was assumed to be flat ($\Omega_{m}+\Omega_{q}=1$). The range of
 scales used for the power spectrum is $10<\ell<2\times10^{5}$.}
 \label{fig:fisher_wd}
\end{figure}

Figure~\ref{fig:fisher_wd} shows the joint constraints on $w$ and
$\Omega_{m}$ which can be derived from the wide and deep wide
surveys. The contours correspond to the 68\% confidence level and
have been marginalized over $h$, $n$ and $\Omega_{b}$. A COBE
prior for the power spectrum normalization $\delta_{h}$ of 7\% rms
(Bunn \& White 1997) was also assumed and marginalized over. The
range of scales considered to evaluate the power spectrum is
$10<\ell<2\times10^{5}$.

Clearly, the wide survey provides stronger constraints than the
deep survey, even though its observing time is 6.4 times shorter.
This follows from the fact that the increased surface density of
resolved galaxies in the deep survey does not compensate for its
smaller area. This can be seen by comparing the error bars for the
power spectrum from each survey (see Figure~\ref{fig:cls_wd} and
the discussion in \S\ref{cls_wd}).

\subsection{Survey Strategy}
\begin{figure}
 \plotone{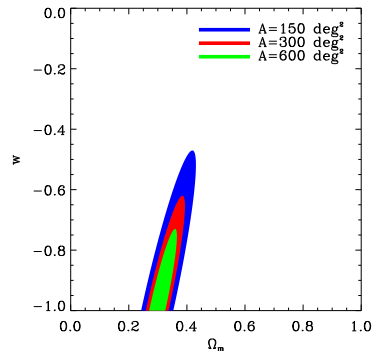}
 \caption{Dependence of the confidence contours on the survey area
 $A$, for a varying survey observing time $t_{\rm tot}$. The depth
 of the survey is fixed to that of the wide SNAP survey (300
 deg$^{2}$, $t_{\rm tot}=5$ months). A survey area of 150 and 600
 deg$^{2}$ would thus require an observing time of 2.5 and 10
 months, respectively. The conventions and marginalizations are as
 described in the caption of  Figure~\ref{fig:fisher_wd}.}
 \label{fig:fisher_a}
\end{figure}

It is instructive to study the dependence of these constraints on
the survey geometry. Figure~\ref{fig:fisher_a} shows how the
constraints on $\Omega_{m}$ and $w$ change as the survey area $A$
is halved or doubled, while the depth of the survey is kept as
that of the wide survey (see parameters for the wide survey in
Table~\ref{tab:nz}). As expected, the contours scale simply as
$A^{-1/2}$ in this case.

\begin{figure}
 \plotone{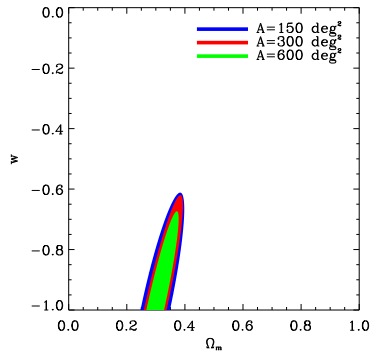}
 \caption{Dependence of the confidence contours on the survey area
 $A$ for a fixed observing time of $t_{\rm tot}=5$ months. This
 corresponds to a trade-off between survey width and depth about
 the nominal SNAP wide survey (300 deg$^{2}$). The sensitivity to
 shear for each exposure time was derived from the image
 simulations described in Paper II.} 
 \label{fig:fisher_a_tconst}
\end{figure}

More realistically, Figure~\ref{fig:fisher_a_tconst} shows the
same contours, but this time keeping the survey observing time
constant to $t_{\rm tot}=5$ months, the allocated time for the
wide survey. This amounts to a trade-off between area and depth
for a fixed observing time.  The survey parameters for each of the
150, 300 and 600 deg$^{2}$ cases are listed in table~\ref{tab:nz},
with entries `Wide$-$', `Wide' and `Wide+', respectively. As can
be seen on the figure, the constraints do not improve as fast as
in the earlier case. Doubling the survey area from 300 to 600
deg$^{2}$, while reducing the depth correspondingly, leads to an
improvement on the precision of $w$ of only about 10\%.

A wider and shallower survey is therefore preferred compared to
the nominal wide survey, but does not provide a substantial
improvement. As explained in Paper I, the shallowness of the
survey is limited by the finite telemetry bandwidth of the
spacecraft, and can not be increased without performing lossy data
compression or modification of the hardware. Moreover, a shallower
survey will limit our ability to measure the redshift dependence
of the lensing power spectrum (see \S\ref{tomography} and
\S\ref{params_tomography} below). These considerations led to the
choice of the baseline survey strategy of the SNAP wide survey
(see Paper I).

\subsection{Tomography}
\label{params_tomography}
As discussed in \S\ref{tomography}, the constraints can be improved by
studying the redshift dependence of the lensing power spectrum. This
can be done by subdividing the galaxy sample into several redshift
bins using photometric redshifts.

\begin{figure}
 \plotone{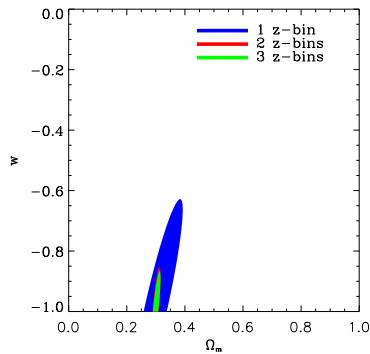}
 \caption{Improvement of the constraints on $w$ and $\Omega_{m}$ from
 the use of tomography. One, two and three redshift bins derived from
 photometric redshifts in the SNAP wide survey (with near-IR detectors)
 are displayed.}
 \label{fig:fisher_z}
\end{figure}

Figure~\ref{fig:fisher_z} shows how the constraints on $w$ and
$\Omega_{m}$ improve when the galaxies in the SNAP wide survey are
split into 2 and 3 redshift bins. The redshift distribution $n(z)$
of each bin are those in the bottom left panel in
Figure~\ref{fig:zbins}.
The parameters for these distributions are listed in
Table~\ref{tab:nz}. The constraints on both $w$ and $\Omega_{m}$
improve by about a factor of 2 in precision when 2 bins are used
instead of 1. The gain from additional bins is not very
significant. This results agrees with the conclusions of Huterer
(2001) and Hu (2001) who considered more generic cases and simpler
redshift distributions. Note that our analysis includes the effect
of photometric redshift errors and of the resulting leakage from
one bin to the other (see overlapping tails in
Figure~\ref{fig:zbins}).

\subsection{Skewness}
\begin{figure}
 \plotone{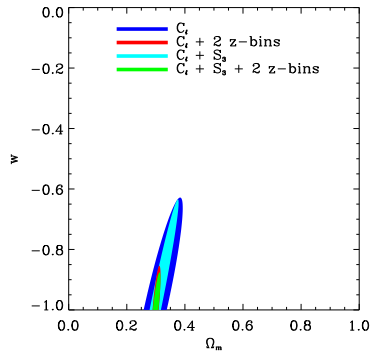}
 \caption{Constraints on $\Omega_{m}$ and $w$ derived from
 combinations of the power spectrum without tomography
 ($C_{\ell}$), the power spectrum with 2 redshift bins, and the
 skewness $S_{3}$. The measurement of $S_{3}$ at the single scale
 $\theta=2'$  is considered.}
 \label{fig:fisher_clskew}
\end{figure}

As discussed in \S\ref{skewness}, another way of improving the
cosmological constraints is to also include a measurement of the
skewness $S_{3}$.  Figure~\ref{fig:fisher_clskew} shows the
contours on the $\Omega_{m}$-$w$ plane corresponding to the use of
the power spectrum with and without tomography (with two redshift
bins) and with and without skewness. As discussed in
\S\ref{skew_onescale}, a measurement of $S_{3}$ at the single
scale of $\theta=2'$ is conservatively considered. The addition of
the skewness improves the precision on $\Omega_{m}$ by a little
less than a factor of 2, but does not appreciably improve the
precision of $w$. The former arises from the well known fact that
a measurement of $S_{3}$ helps to break the degeneracy between the
power spectrum normaliszation and $\Omega_{m}$ (Bernardeau et al.
1997).

The improvements on both $\Omega_{m}$ and $w$ from the inclusion of
the skewness are however overwhelmed by the corresponding improvements
derived from tomography. This shows that tomography is more powerful
than the skewness to study dark energy, at least for conditions
similar to that of the SNAP wide survey. Note that our treatment
of the skewness using the Fisher matrix provides a lower limit for the
parameter errors, since the error of the skewness is non-gaussian
(this is also true for the power spectrum). This conclusion will thus
be a fortiori true for a full non-gaussian treatment of the skewness
error. The combined constraints using both tomography and skewness
are also displayed in Figure~\ref{fig:fisher_clskew}.

\subsection{Comparison with Constraints from Supernovae}
\begin{figure}
 \plotone{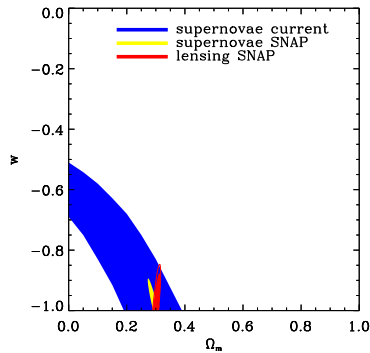}
 \caption{Comparison of the constraints derived from weak lensing and
 from supernovae. The current constraints from 42 supernovae
 (Perlmutter et al. 1999) are also displayed, along with those expected
 from the SNAP deep supernovae survey (Perlmutter et al. 2003). In the
 latter case, the time derivative $w'$ of $w$ was also marginalized
 over. The weak lensing contours assume the use of 2 redshift bins for
 the power spectrum and of the skewness. The filled and unfilled contours
 correspond to the constraints with and without the COBE normalization prior,
 respectively. As before all contours correspond to 68\%
 confidence levels.}
 \label{fig:fisher_sne}
\end{figure}

The results described above show that weak lensing provides powerful
contraints on dark energy which can be compared with those derived
with other methods. Figure~\ref{fig:fisher_sne} compares the
constraints from weak lensing to those from supernovae. The filled weak
lensing contours include tomography (with 2 redshift bins), the
skewness, and the COBE normalization prior. The broad contours
correspond to the current constraints from 42 supernovae (Perlmutter
et al. 1999). The expected constraints derived from supernovae found
in the SNAP deep survey are also shown (Perlmutter et al. 2003). Note
that these authors have marginalised over the time derivative $w'$ of
$w$, and have thus not assumed that $w$ was constant as we have done.
In addition, their constraints, unlike ours, include
uncertainties due to systematics (see discussion on
systematics in \S\ref{conclusion}). As before, all contours
correspond to 68\% confidence levels.

The SNAP weak lensing survey will clearly greatly improve upon the
current supernovae constraints on $w$. It will also yield
constraints which are comparable and somewhat orthogonal to those
derived from the SNAP deep surpernovae survey. Note however that
the SNAP weak lensing survey is obtained from 5 months of
observations rather than 32 months for the deep supernovae survey.

The unfilled contours in figure~\ref{fig:fisher_sne} show the effect
of dropping the COBE prior for the weak lensing constraints. The
precision for $\Omega_{m}$ is hardly affected, but that for $w$ is
degraded by about 50\%. Note that the above conclusions are contingent
on the fact that lensing systematic uncertainties are
subdominant. This will be discussed in the next section.

\section{Conclusions}
\label{conclusion} We have studied the capability of a wide-field
space telescope to measure cosmological parameters with weak
gravitational lensing. For this purpose, we have used the results of
the image simulations described in Paper II to estimate the
sensitivity of the lensing shear for several survey strategies, using
the SNAP mission as a concrete example. By combining the power
spectrum measured in several redshift bins and the skewness of the
convergence field, we find that the SNAP wide survey will provide a
measure $w$ and $\Omega_{m}$ with a 68\%CL uncertainty of
approximately 12\% and 1.5\% respectively.  These errors include
marginalization over other parameters ($h$, $A$, $n$ and $\Omega_{b}$)
using COBE priors for the power spectrum normalization $\delta_{h}$
and assume a flat universe, but neglect systematics (see
discussion below). These constraints are comparable and nearly
orthogonal to those derived from supernovae in the SNAP deep
survey. The constraints on $w$ and $\Omega_{m}$ degrade by a factor of
about 2 in the absence of tomography, but are not affected very much
if the skewness only is dropped.

We also studied how the constraints on these parameters depend on the
survey strategy. We found that, for a fixed observing time of 5
months, they improve slowly if the survey is made wider and
shallower. This, combined with the limits imposed by the spacecraft
telemetry, confirms the choice of the nominal parameters for the SNAP
wide survey.

Note that our analysis relies on a number of assumptions. We first
assumed that systematic errors are sub-dominant compared to
statistical errors. The level of systematics will be greatly reduced
for SNAP, as compared to ground based surveys, thanks to the absence
of the atmospheric seeing and due to the stable thermal orbit of the
spacecraft. This is confirmed by our assessment of the systematics for
the SNAP design described in Paper I. Further instrument and image
simulations are however required to confirm these estimates. In
addition, the SNAP optical and near-IR filter set will allow us to
test and limit the impact of intrinsic galaxy alignements using
photometric redshifts (see Heavens 2001 for a review).

We also assumed that the errors for the power spectrum and skewness
are gaussian, and thus that the fisher matrix provides good estimates
of the errors. We also neglected potential cross talks between the
power spectra in different redshift bins. While these effect are not
expected to have a large influence on our error estimates (see White
\& Hu 2000), these approximations ought to be tested in the future
using N-body simulations.

Another potential limitation arises from the theoretical uncertainties
inherent in the computation of the matter power spectrum and
bispectrum (see discussion in Huterer 2001; van Waerbeke et
al. 2001). Huterer indeed remarked that significant differences exist
between the different available formulae for the non-linear
corrections to the matter power spectrum (eg. Peacock \& Dodds 1996;
Ma et al. 1999) in QCDM models. Larger and more accurate N-body
simulations of QCDM models are needed to improve the accuracy of the
fitting functions and to establish whether the finite accuracy of the
theoretical predictions will be a limitation for the precision reached
by future instruments.

Our work demonstrates that weak lensing is a powerful probe of both
dark matter and dark energy. The complementarity of the constraints
derived from weak lensing and supernovae validates the integration of
both techniques in the science goals for SNAP. A joint analysis of the
constraints which can be derived from weak lensing, supernovae and CMB
anisotropies on both $w$ and its evolution is left to future work.

\acknowledgements We thank Dragan Huterer for useful discussions and
for sharing his results for comparison. We are grateful to Eric Linder
for detailed comments on the manuscript and for fruitful
discussions. The authors have also benefitted from numerous
discussions with the members of the SNAP collaboration. AR was
supported in Cambridge by a PPARC Advanced Fellowship and by a Wolfson
College Research Fellowship.  The authors acknowledge the Sackler fund
in Cambridge for travel allowances.

\end{document}